\documentstyle[12pt]{article}
\begin{document}
\title{GALACTIC ROTATION AND LARGE SCALE STRUCTURES}
\author{B.G. Sidharth\\ Centre for Applicable Mathematics \& Computer Sciences\\
B.M. Birla Science Centre, Hyderabad 500 063}
\date{}
\maketitle
\footnotetext{Email:birlasc@hd1.vsnl.net.in}
\begin{abstract}
On the basis of a recent cosmological model, the puzzle of galactic rotational
velocities at their edges is explained without invoking dark matter. A rationale
for the existence of structures like galaxies and superclusters is also
obtained.
\end{abstract}
\section{Introduction}
In a previous communication\cite{r1} we had described a cosmological scheme
which is consistent with observations and yet does not invoke dark matter. It
is ofcourse well known that the puzzle of galactic rotational velocities can
be explained by the dark matter hypothesis\cite{r2,r3}. Briefly put, using the
well-known relation for rotation under gravitation,
\begin{equation}
\frac{mV^2}{r} = \frac{GMm}{r^2}\label{e1}
\end{equation}
we would expect that from (\ref{e1}) the rotational velocities $V$ at the edges
of galaxies would obey the relation
\begin{equation}
V^2 = \frac{GM}{r}\label{e2}
\end{equation}
where $M$ is the galactic mass, $r$ the radius of the galaxy and $G$, the
gravitational constant. That is the velocities would fall off with increasing
distance from the centre of the galaxy. However, it is observed that these
velocities tend to a constant\cite{r3},
\begin{equation}
V \sim 300 Km/sec\label{e3}
\end{equation}
Alternatively, it is observed that the mass of the universe obeys the law\cite{r4},
\begin{equation}
M \propto R^n, n \approx 1\label{e4}
\end{equation}
These discrepancies can be explained if there is unobserved or unaccounted, that
is missing or dark matter, whose gravitational influence is nevertheless present.
This would also close the universe, that is the expansion would halt and a
collapse would ensue. While no dark matter has yet been discovered, it must
be mentioned that one candidate is a massive neutrino\cite{r5}. Recently, the
Superkamiokande experiments have yielded the first evidence for this\cite{r6},
but it is recognized that this mass, roughly of the order of a billionth that
of the electron is far too small to be the missing or dark matter.\\
On the other hand, latest observations of distant supernovae by different teams
of observers show that the universe would continue to expand for ever\cite{r7,r8}.\\
The cosmological scheme considered in reference \cite{r1} (cf. also ref.\cite{r9})
predicts precisely such a behaviour and moreover, explains (\ref{e4}) without
invoking dark matter. In this scheme, particles, typically pions are
fluctuationally created out of a background ZPF. (Other mysterious, hitherto
empirical relations, like Dirac's large number equations or the inexplicable
Weinberg pion-Hubble constant relation are deduced in this theory).\\
We will now show that in this cosmological scheme, not only the puzzling
galactic rotation relation (\ref{e3})is explained, but also the fact that
structures like galaxies and superclusters would naturally arise.
\section{Galactic Rotation}
We first observe that for a typical galaxy, the mass $'M'$ which is about
$10^{11}$ solar masses, is given by
\begin{equation}
M = Nm = 10^{70}.m\label{e5}
\end{equation}
where $'m'$ is the mass of a typical elementary particle, which in the
literature has been taken to be a pion\cite{r10} and $'N'$ their number.\\
We next observe that the size $'r'$ of a typical galaxy (about a $100000$ light
years) is given by,
\begin{equation}
r = \sqrt{N}l\label{e6}
\end{equation}
where $'l'$ is the pion Compton wavelength and $'N'$ is given in (\ref{e5}).\\
Finally in the cosmological scheme referred to (cf.ref.\cite{r1,r9}), we have,
\begin{equation}
G = a/\sqrt{\bar N}\label{e7}
\end{equation}
where $\bar N \sim 10^{80}$ is the number of pions in the universe and a
$\sim 10^{32}$.\\
Introducing (\ref{e5}), (\ref{e6}) and (\ref{e7}) into (\ref{e2}), we get for
the rotational velocity, because as we go to the edge, the number of particles
$\to N$, the relation (\ref{e3}), consistent with observations.
\section{Large Scale Structures}
It is quite remarkable that equation (\ref{e6}) is true for the universe itself,
as originally pointed out by Eddington, and also for superclusters, as can
be verified. Moreover (\ref{e6}) is a very general relation in the theory of
Brownian motion - it shows that the system under consideration could be
thought of as a collection of these elementary particles in random motion\cite{r11,r12}.
Further, (\ref{e6}) also shows that these structures have an overall flat or
two-dimensional character, which is indeed true\cite{r13}. In particular,
galaxies have vast flat discs and superclusters have a cellular character\cite{r2}.
It must also be pointed out that recent observations do indeed suggest such an
anisotropy\cite{r14}. Finally, the recently discovered neutrino masses are
small, in which case the particles have relativistic speeds, and this is known
to imply the above type of flat structures\cite{r3}.\\
On the other hand, (\ref{e6}) is not valid for stars. At these distance scales,
gravitation is strong enough so that the Brownian approximation (\ref{e6}) is
no longer valid.
\section{Conclusion}
We have thus explained without invoking dark matter, the galactic rotational
relation (\ref{e3}) and also have obtained a rationale for the existence of
structures like galaxies and superclusters.

\end{document}